\documentclass[a4paper,11pt]{article}
\usepackage{jcappub} 
\usepackage{bm}
\usepackage[mathlines]{lineno}
\usepackage{amsfonts}
\usepackage{accents}
\usepackage{latexsym}
\usepackage{braket}
\usepackage{xcolor}
\usepackage{amsmath}
\usepackage{amssymb}
\usepackage{url}
\usepackage{tikz}
\usepackage{pgfplots}
\pgfplotsset{compat=1.18}
\def\nn{\nonumber}
\def\ds{\mathrm{dS}}
\def\dso{\mathrm{dS}_{2}}
\def\dst{\mathrm{dS}_{3}}
\def\dsf{\mathrm{dS}_{4}}

\def\x{\mathbf{x}}
\def\k{\mathbf{k}}
\def\rnd{\partial}
\def\wwp{\mathrm{W}_{\kappa,\gamma}}

\def\wmp{\mathrm{M}_{\kappa,\gamma}}

\def\w{\mathrm{W}}
\def\m{\mathrm{M}}
\def\csch{\mathrm{csch}}

\title{ Massless fermionic current of Schwinger pairs in 3D de Sitter spacetime}

\author[a,*]{Manizheh Botshekananfard} 
\affiliation[a]{Department of Physics, Bo\u{g}azi\c{c}i University, Bebek, 34342, \.{I}stanbul, Turkey}

\author[b]{Cl\'ement Stahl}
\affiliation[b]{Universit\'e de Strasbourg, CNRS, Observatoire astronomique de Strasbourg, UMR 7550, 67000 Strasbourg, Francee} 

\emailAdd{$^{*}$ manizheh.botshekananfard@bogazici.edu.tr}
\emailAdd{clement.stahl@astro.unistra.fr}

\date{\today}

\date{\today}

\begin{abstract}{Pair creation from the vacuum in the presence of a U(1) gauge field in de Sitter  $(\mathrm{dS})$ spacetime provides an important setting for exploring quantum field theory in curved backgrounds. In this work, we investigate massless fermion production and the associated induced current generated by a constant electric field in (1+2)-dimensional  $\mathrm{dS}$ spacetime. Assuming the Bunch--Davies vacuum and employing adiabatic regularization, we derive, for the first time, a finite expression for the induced current of massless fermions in $\mathrm{dS}_{3}$.
The produced fermions generate a net current opposite to the external electric field. In the strong-field regime, the induced current exhibits the expected semiclassical scaling and reproduces the standard Schwinger behavior in the flat-spacetime limit. In the weak-field regime, the current is linear in the electric-field strength. This behavior is characteristic of the fermionic nature of the particles, since the corresponding bosonic case exhibits infrared hyperconductivity, which is absent in our study.
We further show that the induced current remains monotonic throughout the parameter space and, unlike in $\mathrm{dS}_4$, does not exhibit any sign change.
 Our results thus highlight the role of dimensionality and clarify the interplay between spin and infrared physics in $\mathrm{dS}$. This work provides a consistent basis for future investigations including backreaction effects, topologically massive fermions, and time-dependent electromagnetic backgrounds.}
\end{abstract}


\begin{document}
\maketitle
\flushbottom
\section{Introduction}

The instability of the quantum vacuum in the presence of strong electromagnetic fields is one of the fundamental predictions of quantum field theory (QFT). The possibility of field-induced pair creation was first recognized by Sauter \cite{SauterUber:1931fsu}, further developed through the nonlinear effective action of Heisenberg and Euler \cite{Heisenberg:1936whi}, and placed on a fully gauge-invariant footing by Schwinger, whose exact one-loop calculation established vacuum decay as a nonperturbative tunneling process, now known as the Schwinger effect \cite{SchwingerGauge:1951js}. Since then, the Schwinger mechanism has become a paradigmatic example of vacuum instability and has found applications ranging from strong-field Quantum Electrodynamics (QED) to cosmology and astrophysics \cite{Ruffini:2010rrf}.

In curved spacetime, particle production can also occur as a consequence of the dynamical geometry itself. This phenomenon was first demonstrated by Parker in expanding universes \cite{parker1968particle}, and subsequently incorporated into the framework of QFT in curved spacetime through the development of renormalized observables by Davies, Fulling, Christensen, and Bunch \cite{Davies:1977pcw}. In de Sitter (dS) spacetime, these developments culminated in the identification of the de Sitter-invariant Bunch--Davies vacuum \cite{BunchQuantum:1978tcb}, providing the standard framework for studying QFT in an expanding universe.

A central aspect of QFT in curved spacetime is the renormalization of physical observables. Among the available methods, adiabatic regularization has proved particularly successful in cosmological backgrounds, providing a consistent prescription for renormalizing the expectation values of composite operators such as the energy--momentum tensor and induced currents \cite{Birrel:1978ndb}. These techniques form the basis for quantitative studies of vacuum polarization, particle creation, and backreaction in expanding spacetimes.

Motivated by the relevance of dS spacetime to inflationary cosmology, considerable effort has been devoted to understanding the Schwinger effect in the presence of external electromagnetic fields. Pair production, vacuum polarization, induced currents, and the energy--momentum tensor have been investigated for both scalar and fermionic fields in various dimensions using different renormalization prescriptions
\cite{frob2014schwinger,kobayashi2014schwinger,
StahlEckhard:2015csse,stahl2016fermionic,
bavarsad2016scalar,
hayashinaka2016fermionic,
HayashinakaYokoyam:2016,
Sharma:2017rs,
hamil2018schwinger,
THayashinaka:2018thss,
banyeres2018vacuum,
domcke2020chiral,
Gholizadeh:2023ogm,
botshekananfard2020induced,
botshekananfard2025induced,
bastero2026classical}.
One of the most intriguing discoveries in this context is the phenomenon of infrared hyperconductivity (IR-HC). Studies of scalar fields in dS spacetime revealed that, for sufficiently weak electric fields, the induced current may become strongly enhanced, leading to an infrared-dominated regime
\cite{kobayashi2014schwinger,bavarsad2016scalar}. This unexpected behavior attracted considerable attention because it has no analogue in flat spacetime and may have important consequences for cosmological magnetogenesis and the backreaction of produced particles \cite{kobayashi2014schwinger,stahl2019schwinger}, but its physical interpretation is still debated \cite{banyeres2018vacuum,bastero2026classical}. Interestingly, the situation appears to be different for fermions. Previous analyses of fermionic induced currents in $\mathrm{dS}_4$ found no evidence of IR-HC
\cite{hayashinaka2016fermionic}. Instead, the renormalized current remains well behaved in the infrared, suggesting that the low-energy response depends sensitively on the spin and statistics of the quantum field. Whether this conclusion also holds for massless fermions in odd-dimensional dS spacetime remains unknown.

Fermionic current in ${\rm{dS}}_4$ may change sign \cite{hayashinaka2016fermionic}. Its physical interpretation remains under debate and has been linked to the interplay between vacuum polarization, pair production, and cosmological expansion \cite{HayashinakaYokoyam:2016, THayashinaka:2018thss}.
Lower-dimensional QFTs provide an ideal laboratory for addressing this question. Besides being technically more tractable, they often exhibit enhanced infrared and topological effects that are absent in higher dimensions. Moreover, previous studies have revealed interesting correspondences between bosonic and fermionic induced currents in $\mathrm{dS}_2$ \cite{stahl2016fermionic}, while recent investigations of vacuum polarization and the energy--momentum tensor in lower-dimensional QED in dS have further emphasized the role of dimensionality in determining the vacuum response
\cite{botshekananfard2020induced,botshekananfard2025induced}. Nevertheless, an analytic study of the induced current generated by massless fermions in $\mathrm{dS}_3$ is still lacking.

The present work fills this gap. We investigate the induced current generated by massless fermions in $\mathrm{dS}_3$ in the presence of a constant electric field. Employing the Bunch--Davies vacuum together with adiabatic regularization, we derive an analytic expression for the renormalized current and investigate its asymptotic behavior both in the strong- and weak-field regimes. We show that the induced current remains finite and analytic in the weak-field limit, exhibiting no IR-HC. Furthermore, the current remains positive throughout the parameter space considered, in contrast to the sign-changing behavior reported for scalar fields. Our analysis therefore clarifies the role of dimensionality and spin in determining the infrared response of quantum fields in dS.

The massless limit considered here is of particular interest because it allows for an analytic treatment of the induced current while isolating the effects of spacetime curvature and the external electric field from mass-dependent corrections. It also provides a natural starting point for future investigations of massive fermions, where both parity-conserving and parity-violating mass terms may introduce qualitatively new features.

The paper is organized as follows. In Sec.~\ref{sec:dS}, we briefly review the Dirac equation in dS spacetime and construct the corresponding spinor mode functions. In Sec.~\ref{sec:Current}, we derive the vacuum expectation value of the induced current. Section~\ref{sec:adiabatic regularization} presents the adiabatic regularization procedure and the renormalized current. In Sec.~\ref{sec:result}, we analyze the asymptotic behavior of the current in the strong- and weak-field limits. Finally, Sec.~\ref{sec:conclusion} contains our conclusions. Technical details are collected in Appendices~\ref{app:whittaker} and~\ref{computation}.

\section{Solutions of the Dirac equation in dS spacetime}
\label{sec:dS}
To study the vacuum expectation value of the current operator of a charged Dirac field coupled to a constant electric field background in $\ds$, we will compute the fermion field modes. Since the field operators are constructed from these mode functions, the Dirac equation in the presence of a constant electric field background in dS spacetime must first be solved.
\subsection{Basics of spinors in dS spacetime}
\label{sec:basic}
The $\dst$ spacetime metric can be read from the line element in the half of $\dst$ manifold
\begin{align}\label{linerw}
ds^{2}&=dt^{2}-e^{2Ht}d\x^{2}, & t&\in(-\infty,\infty), & \x&\in\mathbb{R}^{2},
\end{align}
which is the line element of an expanding universe with increasing proper time $t$ and $H$ is the Hubble constant. In terms
of the conformal time
\begin{align}\label{time}
\tau&=-\frac{1}{H}e^{-Ht}, & \tau&\in(-\infty,0),
\end{align}
the line element~(\ref{linerw}) takes the form
\begin{align}\label{ds}
ds^{2}&=\Omega^{2}(\tau)(d\tau^{2}-d\x^{2}), & \Omega(\tau)&=-\frac{1}{\tau H},
\end{align}
revealing that this portion of dS is conformal to a portion of Minkowski spacetime \cite{BirrellDavies1984}.
\par
The anticommutation relations are satisfied by spacetime-dependent Dirac $\gamma$-matrices in curved spacetime with metric $g_{\mu\nu}$:
\begin{equation}\label{gammag}
\big\{{\Gamma^{\mu}(x),\Gamma^{\nu}}(x)\big\}=2g^{\mu\nu}(x).
\end{equation}
An explicit representation of the $\Gamma^{\mu}(x)$ matrices in terms of Minkowski spacetime $\gamma^{a}$ matrices is obtained
by introducing tetrad fields
\begin{equation}\label{tetrad}
g_{\mu\nu}(x)=e^{a}{\,}_{\mu}(x)e^{b}{\,}_{\nu}(x)\eta_{ab},
\end{equation}
where $\eta_{ab}$ is the Minkowski spacetime metric. Then, $\Gamma^{\mu}(x)$ matrices can be written as follows
\begin{equation}\label{gammax}
\Gamma^{\mu}(x)=g^{\mu\nu}\eta_{ab}e^{a}{\,}_{\nu}(x)\gamma^{b}.
\end{equation}
In (2+1)-dimensional spacetime, Dirac spinors are two-component objects, in contrast to the four-component spinors encountered in (3+1) dimensions. This distinction is important for interpreting the induced current and comparing our results with those in higher-dimensional  $\mathrm{dS}$ spacetimes.
The spinorial affine connections $B_{\mu}(x)$ are matrices defined the vanishing of the covariant derivative of the
$\Gamma^{\mu}$ matrices
\begin{equation}\label{gammad}
\nabla_{\mu}\Gamma_{\nu}=\partial_{\mu}\Gamma_{\nu}-\Gamma_{\mu\nu}^{\lambda}\Gamma_{\lambda}
+B_{\mu}\Gamma_{\nu}-\Gamma_{\nu}B_{\mu}=0,
\end{equation}
where $\Gamma_{\mu\nu}^{\lambda}$ are Christoffel connections. It can be shown \cite{parker2009quantum} that the spinorial affine
connections $B_{\mu}(x)$ satisfy Eq.~(\ref{gammad}) if
\begin{equation}\label{affine}
B_{\mu}(x)=\frac{1}{8}e^{b}{\,}_{\kappa}g^{\kappa\nu}\big(\Gamma_{\mu\nu}^{\lambda}e^{a}{\,}_{\lambda}
-\partial_{\mu}e^{a}{\,}_{\nu}\big)\big[\gamma_{a},\gamma_{b}\big].
\end{equation}
Consequently, the covariant derivative of a Dirac spinor field is
\begin{equation}\label{nabla}
\nabla_{\mu}\psi=(\partial_{\mu}+B_{\mu})\psi.
\end{equation}
The conformal dS spacetime metric on the Poincar\'{e} patch can be read from Eq.~\eqref{ds}
\begin{equation}\label{metric}
g_{\mu\nu}(\tau)=\Omega^{2}(\tau)\eta_{\mu\nu}.
\end{equation}
By using Eq.~(\ref{tetrad}), the tetrad fields are given by in the diagonal gauge as
\begin{equation}\label{tetrads}
e^{a}{\,}_{\mu}(\tau)=\Omega(\tau)\delta^{a}_{\mu}.
\end{equation}
As a result of Eqs.~(\ref{tetrad}) and~(\ref{tetrads}), the explicit form of the $\Gamma^{\mu}(x)$ matrices is obtained
\begin{equation}\label{gammads}
\Gamma^{\mu}(\tau)=\Omega^{-1}(\tau)\gamma^{\mu}.
\end{equation}
We chose Minkowski spacetime $\gamma$-matrices in the Weyl representation
\begin{align}\label{weyl}
\gamma^{0}&=\sigma_{1}, & \gamma^{1}&=i\sigma_{2}, & \gamma^{2}&=i\sigma_{3},
\end{align}
where $\sigma_{1},{\,}\sigma_{2}$ and $\sigma_{3}$ are the Pauli matrices. Substituting the tetrad fields~(\ref{tetrads}),
Christoffel connections associated with metric~(\ref{metric})
and $\gamma$-matrices~(\ref{weyl}), into  Eq.~(\ref{affine}) the spinorial affine connections $B_{\mu}(x)$ are obtained
\begin{equation}\label{affineds}
B_{\mu}(\tau)=\frac{1}{2}H\Omega(\tau)\big[\sigma_{2}\delta_{\mu,2}-\sigma_{3}\delta_{\mu,1}\big].
\end{equation}

\subsection{The Dirac equation} \label{sec:Dirac}
In order to obtain the solutions of the Dirac equation in presence of a constant electric field background, we consider the
action of QED in the $\dst$ spacetime
\begin{equation}\label{action}
S=\int d^{3}x\sqrt{|g|}\Big\{\frac{i}{2}\overline{\psi}\Gamma^{\mu}\big(\nabla_{\mu}+ieA_{\mu}\big)\psi
-\frac{i}{2}\big[\big(\nabla_{\mu}-ieA_{\mu}\big)\overline{\psi}\,\big]\Gamma^{\mu}\psi
-\frac{1}{4}F_{\mu\nu}F^{\mu\nu}\Big\},
\end{equation}
where $\psi(x)$ is a two-component spinor field  without mass and electric charge $e$. The metric of $\dst$ spacetime is given
by Eq.~(\ref{metric}) and $|g|$ is absolute value
of its determinant. Matrices $\Gamma^{\mu}$ read off from Eqs.~(\ref{gammads}) and~(\ref{weyl}). The covariant derivatives in $\dst$ are determined by Eqs.~(\ref{nabla}) and~(\ref{affineds}). The adjoint spinor field is defined as
\begin{equation}\label{psibar}
\overline{\psi}(x):=\psi^{\dag}(x)\gamma^{0},
\end{equation}
where $\gamma^{0}$ is given by Eq.~(\ref{weyl}). We will take the vector potential describing a constant electric field
background to be
\begin{equation}\label{vector}
A_{\mu}(\tau)=-\frac{E}{H^{2}\tau}\delta_{\mu,1},
\end{equation}
where $E$ is a constant. Then, the only non-zero components of the electromagnetic field strength tensor are
\begin{eqnarray}\label{tensor}
F_{01}&=&-F_{10} \nn \\
&=&\rnd_{0}A_{1}-\rnd_{1}A_{0}=E\Omega^{2}(\tau).
\end{eqnarray}
We choose the electric field to point along the x-direction for convenience. This choice is arbitrary and is related to any other spatial direction by a spatial rotation.
The equations of motion for the spinor field $\psi(x)$ are obtained by varying the action~(\ref{action}) under an infinitesimal displacement
$\overline{\psi}(x)\rightarrow\overline{\psi}(x)+\delta\overline{\psi}(x)$. We get
\begin{equation}\label{diraceq}
\big[i\Gamma^{\mu}\big(\nabla_{\mu}+ieA_{\mu}\big)\big]\psi=0.
\end{equation}
Substituting Eqs.~(\ref{gammads},\ref{weyl},\ref{affineds},\ref{vector}) into the Dirac equation~(\ref{diraceq}) yields
\begin{equation}\label{psieq}
\Big[i\sigma_{1}\rnd_{0}-\sigma_{2}\Big(\rnd_{1}+i\frac{eE}{H}\Omega\Big)-\sigma_{3}\rnd_{2}
+iH\Omega\sigma_{1}\Omega\Big]\psi=0.
\end{equation}
If we define
\begin{equation}\label{tilde}
\tilde{\psi}(x):=\Omega(\tau)\psi(x).
\end{equation}
One can show that Eq.~(\ref{psieq}) leads to
\begin{equation}\label{tildeq}
\Big[i\sigma_{1}\rnd_{0}-\sigma_{2}\Big(\rnd_{1}+i\frac{eE}{H}\Omega\Big)
-\sigma_{3}\rnd_{2}\Omega\Big]\tilde{\psi}=0.
\end{equation}
Multiplying both sides of Eq.~(\ref{tildeq})
\begin{equation}\label{Multiply}
\Big[i\sigma_{1}\rnd_{0}-\sigma_{2}\Big(\rnd_{1}+i\frac{eE}{H}\Omega\Big)
-\sigma_{3}\rnd_{2}\Omega\Big]\Big[i\sigma_{1}\rnd_{0}-\sigma_{2}
\Big(\rnd_{1}+i\frac{eE}{H}\Omega\Big)-\sigma_{3}\rnd_{2}\Omega\Big]\tilde{\psi}=0,
\end{equation}
after some algebra, we obtain
\begin{equation}\label{algebra}
\Big[\rnd_{0}^{2}-\Big(\rnd_{1}+i\frac{eE}{H}\Omega\Big)^{2}-\rnd_{2}^{2}-ieE\Omega^{2}\sigma_{3}\Big]\tilde{\psi}=0.
\end{equation}
Now, we consider $\tilde{\psi}(x)$ as a two-component spinor field
\begin{equation}\label{spinor}
\tilde{\psi}(x)=
\left[
\begin{array}{c}
\tilde{\psi}_{1}(x) \\
\tilde{\psi}_{2}(x) \\
\end{array}
\right],
\end{equation}
then Eq.~(\ref{algebra}) leads to
\begin{equation}\label{index}
\Big[\rnd_{0}^{2}-\rnd_{1}^{2}-\rnd_{2}^{2}-\frac{2ieE}{H}\Omega\rnd_{1}
+\frac{e^{2}E^{2}}{H^{2}}\Omega^{2}+i(-1)^{s}\lambda H^{2}\Omega^{2}\Big]\tilde{\psi}_{s}=0,
\end{equation}
with $s=1,2$. Based on the invariance of Eq.~(\ref{index}) under the translation along the spatial direction, let
\begin{equation}\label{hatpm}
\tilde{\psi}_{s}^{\pm}(\tau,\x)=e^{\pm i\k\cdot\x}\chi_{s}^{\pm}(\tau),
\end{equation}
here the superscript $\pm$ denotes the positive and negative frequency solutions respectively. Substituting~(\ref{hatpm}) into Eq.~(\ref{index}) leads to
\begin{equation}\label{chieq}
\frac{d^{2}\chi_{s}^{\pm}(z_{\pm})}{dz_{\pm}^{2}}+\Big(-\frac{1}{4}+\frac{\kappa}{z_{\pm}}
+\frac{1/4-\gamma_{s}^{2}}{z_{\pm}^{2}}\Big)\chi_{s}^{\pm}(z_{\pm})=0,
\end{equation}
where the variables $z_{+}$ and $z_{-}$ are defined as
\begin{align}\label{zpm}
z_{+}&:=+2ik\tau, & z_{-}&:=-2ik\tau=e^{i\pi}z_{+},
\end{align}
and $k=+|\k|$. In terms of the dimensionless parameters
\begin{align}\label{lambda}
\lambda&:=\frac{eE}{H^{2}}, & r&:=\frac{k_{x}}{k},
\end{align}
Note that some other groups use the opposite sign convention for $\lambda$, which reverses the sign of quantities expressed in terms of $\lambda$ such as the induced current $J$ that we define in Eq.~\eqref{opcurrent}.
The coefficients $\kappa$, $\gamma$ and $\gamma_{s}$ are defined by
\begin{eqnarray}
\kappa&=&i\lambda r, \label{kappa} \\
\gamma&=&i\lambda \label{gamma}, \\
\gamma_{s}&=&\gamma-\frac{(-1)^{s}}{2}. \label{gammas}
\end{eqnarray}
The differential equation~(\ref{chieq}) is nothing but the Whittaker equation (cf.~Appendix \ref{app:whittaker}). Its most general solution, following the conventions of Ref.~\cite{frank2010nist}, can be written as
\begin{equation}\label{chipm}
\chi_{s}^{\pm}(z_{\pm})=C_{1}\w_{\pm\kappa,\pm\gamma_{s}}(z_{\pm})+C_{2}\m_{\pm\kappa,\pm\gamma_{s}}(z_{\pm}),
\end{equation}
where $C_{1},\,C_{2}$ are arbitrary constant coefficients.

\subsection{Mode functions} \label{sec:mode}
To determine the vacuum state of this QFT, we need the mode functions that define the creation and annihilation operators. This vacuum is defined by specifying the asymptotic form of the mode functions \cite{BirrellDavies1984,parker2009quantum}. In the limit $\tau \to -\infty$,
we impose the initial condition: $\chi_{s}^{\pm}(\tau)$  (Eq.~(\ref{chipm})) asymptotically takes
the form $\chi_{s}^{\pm}(\tau)\sim e^{\mp ik\tau}$. A comparison with Minkowski spacetime mode functions shows that the functions
$\chi_{s}^{+}(\tau)$ and $\chi_{s}^{-}(\tau)$
are positive and negative frequency mode functions, respectively. By virtue of the asymptotic expansions of the Whittaker functions $\w_{\pm\kappa,\pm\gamma_{s}}(z_{\pm})$ and $\m_{\pm\kappa,\pm\gamma_{s}}(z_{\pm})$ as $|z_{\pm}|\rightarrow\infty$, that are shown in Eqs.~(\ref{win}) and~(\ref{min}) respectively, the normalized positive-frequency mode function with the desired asymptotic form is
\begin{align}\label{uin}
U_{in}(x)&=\Omega^{-1}(\tau)e^{+i\k\cdot\x}\chi_{in}^{+}, &
\chi_{in}^{+}&= e^{\frac{i\kappa\pi}{2}}\left[
\begin{array}{c}
\sqrt{\frac{\gamma+\kappa}{2\gamma}} \w_{\kappa,\gamma-\frac{1}{2}}(z_{+}) \\
\sqrt{\frac{\gamma-\kappa}{2\gamma}}\w_{\kappa,\gamma+\frac{1}{2}}(z_{+}) \\
\end{array}
\right],
\end{align}
and the normalized negative-frequency mode function with the desired asymptotic form is given by
\begin{align}\label{vin}
V_{in}(x)&=\Omega^{-1}(\tau)e^{-i\k\cdot\x}\chi_{in}^{-}, &
\chi_{in}^{-}&=e^{\frac{i\kappa\pi}{2}}\left[
\begin{array}{c}
\sqrt{\frac{\gamma-\kappa}{2\gamma}} \w_{-\kappa,\gamma-\frac{1}{2}}(z_{-}) \\
\sqrt{\frac{\gamma+\kappa}{2\gamma}} \w_{-\kappa,\gamma+\frac{1}{2}}(z_{-}) \\
\end{array}
\right], 
\end{align}
where the subscript in denotes that these mode functions have the required asymptotic form at early times, with the corresponding vacuum state referred to as the in-vacuum.
\par

Similarly, to determine the mode functions in the late-time limit ($t\rightarrow+\infty$), we examine the asymptotic form of the functions $\chi_{s}^{\pm}(\tau)$ as $\tau\rightarrow0^{-}$. The asymptotic expansions of the Whittaker functions $\m_{\pm\kappa,\pm\gamma_{s}}(z_{\pm})$ and $\w_{\pm\kappa,\pm\gamma_{s}}(z_{\pm})$ as $|z|\rightarrow0$ are given in Eqs.~(\ref{mout}) and~(\ref{wout}). These asymptotic expansions require the most general solutions~(\ref{chipm}) to contain only the Whittaker functions of $\m$-type, which lead to the desired asymptotic form $\chi_{s}^{\pm}(\tau)\sim e^{\mp i|\gamma|t}$. A direct comparison with the Minkowski spacetime mode functions shows that the functions $\chi_{s}^{+}(\tau)$ and $\chi_{s}^{-}(\tau)$ correspond to the positive- and negative-frequency mode functions, respectively. Hence, the normalized positive-frequency mode function is
\begin{align}\label{uout}
U_{out}(x)&=\Omega^{-1}(\tau)e^{+i\k\cdot\x}\chi_{out}^{+}, &
\chi_{out}^{+}&= e^{i\frac{\gamma\pi}{2}}\left[
\begin{array}{c}
\m_{\kappa,\gamma-\frac{1}{2}}(z_{+}) \\
\sqrt{\frac{\gamma^{2}-\kappa^{2}}{4\gamma^{2}(1-4\gamma^{2})}}\m_{\kappa,\gamma+\frac{1}{2}}(z_{+}) \\
\end{array}
\right], 
\end{align}
and the normalized negative frequency mode function with the desired asymptotic form is
\begin{align}\label{vout}
V_{out}(x)&=\Omega^{-1}(\tau)e^{-i\k\cdot\x}\chi_{out}^{-}, &
\chi_{out}^{-}&=e^{i\frac{\gamma\pi}{2}}\left[
\begin{array}{c}
\sqrt{\frac{\gamma^{2}-\kappa^{2}}{4\gamma^{2}(1-4\gamma^{2})}}\m_{-\kappa,-\gamma+\frac{1}{2}}(z_{-}) \\
\m_{-\kappa,-\gamma-\frac{1}{2}}(z_{-}) \\
\end{array}
\right].
\end{align}
We can now construct the spinor field operator by expanding terms of the $\{U_{in \boldsymbol{k}},V_{in \boldsymbol{k}}\}$ given by Eqs.~(\ref{uin}) and~(\ref{vin})
\begin{equation}\label{psi}
\psi(x)=\int \frac{d^{2}k}{(2\pi)^2}\Big[a_{in\boldsymbol{k}}U_{in,\boldsymbol{k}}(x) +b^{\dag}_{in\boldsymbol{k}}V_{in,\boldsymbol{k}}(x)\Big],
\end{equation}
where $a_{in\boldsymbol{k}}$ is the annihilation operator for particles described by the mode function $U_{in,\boldsymbol{k}}$, and $b^{\dag}_{in\boldsymbol{k}}$ is the creation operator for  antiparticles described by the mode function $V_{in \boldsymbol{k}}$.
The creation and annihilation operators follow the anticommutation relations 
\begin{equation}\label{anticommutation1}
\{a_{in \boldsymbol{k}},a^{\dag}_{in \boldsymbol{k'}}\}=\{b_{in \boldsymbol{k}},b^{\dag}_{in \boldsymbol{k'}}\}=(2\pi)^{2}\delta^{2}(\boldsymbol{k}-\boldsymbol{k'})
\end{equation}
\begin{equation}\label{anticommutation2}
\{a_{in \boldsymbol{k}},b_{in \boldsymbol{k'}}\}=\{b^{\dag}_{in \boldsymbol{k}},a^{\dag}_{in \boldsymbol{k'}}\}=0
\end{equation}
The Bunch-Davies vaccum is defined as
\begin{equation}\label{vaccume}
a_{in\boldsymbol{k}}|0\rangle_{in}=0\;\;\;\;\;\;\;\;\;\;\;\;\;\;\;\;\;\;\;,\forall \boldsymbol{k}.
\end{equation}

\section{In-vacuum state expectation of the current operator} \label{sec:Current}
The fermionic current operator is defined as follows 
\begin{align}\label{opcurrent}
J^{\mu}=-\frac{e}{2}\langle 0| [\bar{\psi}(x),\gamma^{\mu}\psi(x)]|0\rangle,
\end{align}
it is conserved, i.e., $\nabla_{\mu}J^{\mu}=0$ \cite{BirrellDavies1984}. After some algebra, Eq.~(\ref{opcurrent}) yields the in-vacuum state expectation value of the current operator for the Dirac field coupled to the uniform electric field background in a $\dst$:
\begin{eqnarray}\label{currentJK}
\langle J^{1}\rangle_{in} &=&\frac{e}{2}\Omega^{-3}(\tau)   \int \frac{d^{2}k}{(2\pi)^{2}} e^{i\kappa\pi}\bigg[(1+r)\big|W_{\kappa,i \lambda-\frac{1}{2}}(z_{+})\big|^{2}-(1-r)\big|W_{\kappa,i \lambda+\frac{1}{2}}(z_{+})\big|^{2}\bigg].
\end{eqnarray}
To calculate the vacuum expectation value of the current operator
(\ref{opcurrent}), we choose the in-vacuum state since this state is Hadamard \cite{frob2014schwinger,garriga1994pair}. Hence, the
expectation value has a UV behavior similar to that of flat spacetime. After transforming to polar coordinates, performing a change of variables, and substituting the explicit expressions, the integral (\ref{currentJK}) can be written as
\begin{eqnarray}\label{currentJP}
\langle J^{1}\rangle_{in} &=&\frac{e}{(2\pi)^{2}}H^{2}\Omega^{-1}(\tau) \int_{-1}^{+1}\frac{dr}{\sqrt{1-r^{2}}} \int_{0}^{\Lambda}  dp p e^{-\pi \lambda r}  \nn \\
&\bigg[&(1+r) \big|W_{i\lambda r,i \lambda-\frac{1}{2}}(-2ip)\big|^{2}-(1-r)\big|W_{i\lambda r, i\lambda+\frac{1}{2}}(-2ip)\big|^{2}\bigg],
\end{eqnarray}
where $\Lambda=-K \tau$,  in which $K$ is an upper cutoff on momentum $k$. The momentum for the spinor is defined as,
\begin{align}\label{p}
p=-k \tau.
\end{align}
The detailed calculation of these integrals can be found in Appendix \ref{computation}, and here we show only the final result for this Section:
\begin{eqnarray}\label{currentfinal}
\langle J^{1}\rangle_{in} &=&\frac{e}{2\pi^{2}}H^{2}\Omega^{-1}(\tau)\Bigg[\frac{\pi}{2} \lambda\Lambda-\frac{3}{8}i\pi\lambda-\frac{\pi}{4}(1+\lambda^{2})\coth(2\pi\lambda) \nn \\
&+&\frac{\pi}{4}(1-2\lambda^{2}) I_{0}(2\pi\lambda)\csch(2\pi\lambda) 
+\frac{3}{4}(1+i\pi)\lambda I_{1}(2\pi\lambda) \csch(2\pi\lambda) \nn \\
&+&\frac{3}{4}\frac{\lambda^{2}}{\sinh(2\pi\lambda)} \int_{-1}^{+1}dr r\sqrt{1-r^{2}}\nn \\ 
& \times & \bigg\{(e^{-2\pi\lambda r}-e^{-2\pi\lambda})\Psi(i\lambda-i \lambda r)+(e^{2\pi\lambda}-e^{-2\pi \lambda r})\Psi(-i\lambda-i \lambda r)\bigg\}\Bigg],
\end{eqnarray}
where $\Psi$ is the digamma function,  which is defined by the first derivative of the logarithm of the Gamma function \cite{frank2010nist}. $I_{0}$ and $I_{1}$ are the modified Bessel functions.

\section{Adiabatic regularization} \label{sec:adiabatic regularization}
As the in-vacuum state is Hadamard \cite{frob2014schwinger}, the expectation value of the fermionic current has a UV behavior similar to the flat spacetime. To remove the ultraviolet divergence from the expectation values given by Eq.~\eqref{currentfinal}, we apply the adiabatic subtraction scheme. This method has been used to regularize the induced fermionic current in $\dso$ \cite{stahl2016fermionic} and  $\dsf$ \cite{hayashinaka2016fermionic}, and induced fermionic energy-momentum tensor in $\dso$ \cite{botshekananfard2020induced}. We follow a process similar to that used in Ref.~\cite{stahl2016fermionic, botshekananfard2020induced}. We start from the Dirac equation for the spinor mode of positive frequency which is obtained from the action~\eqref{action}
\begin{align}\label{regularmod}
\big[\partial_{0}^{2}+\omega^{2}(\tau)+i(-1)^{s}\sigma(\tau)\big]\mathcal{U}_{s}(\tau) =0, 
\end{align}
where,
\begin{align}\label{omegasigma} 
\omega(\tau) &=+\sqrt{k^{2}+2\lambda H\Omega k r+\lambda^{2}\Omega^{2}H^{2}}, 
 &\sigma(\tau) &=\lambda\Omega^{2}H^{2}.
\end{align}
In this case, $s=1,2$  represents the spinor component index. The solution of equation Eq.~\eqref{regularmod} is considered a WKB solution
\begin{align}\label{wkb} 
\mathcal{U}_{s}(\tau)=N_{s}\exp\left\{-i\int^{\tau} d\tau' \Big(X_{s}(\tau')+i Y_{s}(\tau')\Big)\right\}d\tau',
\end{align}
where $N_{s}$ is a normalized coefficient, and $X_{s}$ and $Y_{s}$ are real functions to be determined by the equation of motion. Substituting it into Eq.~\eqref{regularmod}, one finds
\begin{eqnarray}\label{XsYs}
X_{s}^{2}(\tau)-Y_{s}^{2}(\tau)-\dot{Y_{s}}(\tau)-\omega^{2}(\tau) &=&0, \label{XsYs1}  \\
\dot{X_{s}}(\tau)+2X_{s}(\tau)Y_{s}(\tau)-(-1)^{s}\sigma(\tau) &=& 0, \label{XsYs2}
\end{eqnarray}
where dots refer to the partial derivative with respect to the conformal time. The solution of  Eqs.~\eqref{XsYs1} and \eqref{XsYs2} is given by
\begin{align}\label{wkb2} 
\mathcal{U}_{s}(\tau)=\frac{N_{s}}{ \sqrt{X_{s}(\tau)}}\exp\left\{-i\int^{\tau} \Big(X_{s}(\tau')+i(-1)^{s}\frac{\sigma(\tau')}{2X_{s}(\tau')}\Big)d\tau'\right\}.
\end{align}
 In adiabatic conditions, we assume that background gravitational and electromagnetic fields vary slowly. Hence, the adiabatic expansion is  a power series in terms of degrees of metric derivatives,  we ignore the derivative terms in Eqs.~\eqref{XsYs1} and ~\eqref{XsYs2} at the zeroth order, and obtain
\begin{align}\label{lineconXY}
Y_{s}(\tau)&=\frac{(-1)^{s}\sigma}{2X_{s}(\tau)}, & X_{s}(\tau)&=\omega(\tau)\big(1+\mathcal{O}(\omega^{-2} )\big),
\end{align}
In this step, we define the auxiliary function $F_{k}(\tau)=\frac{\omega(\tau)\dot{\omega}(\tau)}{\sigma(\tau)}$. It can be shown that this function satisfies the following relation
\begin{align}\label{Fk}
\frac{d}{d\tau}\bigg[\log\big(\omega(\tau)+F_{k}(\tau)\big)\bigg]=\frac{\sigma(\tau)}{\omega(\tau)}.
\end{align}
Inserting Eqs.~\eqref{lineconXY} and \eqref{Fk} into Eq.~\eqref{wkb2}, we find that $\mathcal{U}_{s}(\tau)$ can be expressed as
\begin{align}\label{Us}
\mathcal{U}_{s}(\tau)=\frac{N_{s}}{\sqrt{\omega(\tau)}} \bigg(\frac{\omega(\tau)\big[\sigma(\tau)+\dot{\omega}(\tau)\big]}{\sigma(\tau)}\bigg)^{\frac{(-1)^{s}}{2}} \exp\left\{-i\int^{\tau} d\tau' \omega(\tau') \right\}.
\end{align}
To calculate the counterterm, we need the ratio $\mathcal{K}$, which is independent of the normalization coefficient. Substituting Eq.~\eqref{Us} into Eq.~\eqref{tildeq}, we find  
\begin{equation}\label{mathcalk}
 \frac{\mathcal{U}_{2}}{\mathcal{U}_{1}}=\frac{i\big [\omega(\tau)+ k r+ \lambda H\Omega (\tau)\big]}{k \sqrt{1-r^{2}}}.
\end{equation}
 We constrain the components of the adiabatic mode spinor \eqref{wkb} to obey the normalization condition
\begin{equation}\label{normalization}
\big|\mathcal{U}_{1}(\tau)\big|^{2}+\big|\mathcal{U}_{2}(\tau)\big|^{2}=1.
\end{equation}

Following the adiabatic regularization method, the counterterm is constructed from the expectation values of the fermionic current \eqref{opcurrent}, evaluated in the adiabatic vacuum defined by the mode functions \eqref{Us}. These counterterms are given by
\begin{equation}\label{JU}
\langle J^{1}\rangle_{A}
=
e\,\Omega^{-3}(\tau)
\int \frac{d^{2}k}{(2\pi)^{2}}
\left(
\mathcal{U}_{1}\mathcal{U}_{1}^{*}- \mathcal{U}_{2}\mathcal{U}_{2}^{*}
\right).
\end{equation}
By using Eq.~\eqref{mathcalk}, one can rewrite Eq.~\eqref{JU} as
\begin{equation}
\langle J^{1}\rangle_{A}
=
\frac{e\,\Omega^{-1}(\tau)H^{2}}{2\pi^{2}}
\int_{-1}^{+1}\frac{dr}{\sqrt{1-r^{2}}}
\int_{0}^{\Lambda} p\,d p\,
\left(
\frac{
1-\left| \frac{\mathcal{U}_{2}}{\mathcal{U}_{1}} \frac{\mathcal{U}_{2}^{*}}{\mathcal{U}_{1}^{*}}\right|
}{
1+\left| \frac{\mathcal{U}_{2}}{\mathcal{U}_{1}} \frac{\mathcal{U}_{2}^{*}}{\mathcal{U}_{1}^{*}}\right|
}
\right),
\end{equation}
we finally obtain
\begin{eqnarray}\label{countertermcurrent}
\langle J^{1}\rangle_{A} &=&\frac{e}{4\pi} H^{2}\Omega^{-1}(\tau) \lambda\Lambda.
\end{eqnarray}

The adiabatic counterterm in both bosonic and fermionic theories is fully determined by the UV-divergent part of the adiabatic expansion and is linear in the cutoff $\Lambda$. As in the bosonic case, no finite $\Lambda$-independent contribution appears.
Although the present fermionic theory is massless, whereas the bosonic case considered in Ref.~\cite{bavarsad2016scalar} involves a nonzero mass, the overall renormalization procedure remains the same. In both cases, the counterterm removes only the ultraviolet divergence, leaving the finite part unchanged.
The main differences arise from the spinor nature of the fermionic field. The spinor mode decomposition, together with the anticommutation relations of fermionic operators, modifies the structure of the mode sums and the adiabatic expansion compared to the scalar case. Consequently, some terms appear with different numerical prefactors and relative signs than in the bosonic calculation. These differences reflect the underlying spin-statistics properties of the field rather than any change in the subtraction prescription.

Subtracting the second adiabatic order counterterms, given by Eqs.~\eqref{countertermcurrent}  from the corresponding
original expressions for the in-vacuum expectation values which are given by Eqs.~\eqref{currentfinal}, leads to the regularized induced current
\begin{eqnarray}\label{regularizedcurrent}
J& =&\langle J^{1}\rangle_{in}-\langle J^{1}\rangle_{A} \nn \\
 &=&\frac{e}{2\pi^{2}}H^{2}\Omega^{-1}(\tau)\Bigg[-\frac{3}{8}i\pi\lambda-\frac{\pi}{4}(1+\lambda^{2})\coth(2\pi\lambda) \nn \\
&+&\frac{\pi}{4}(1-2\lambda^{2}) I_{0}(2\pi\lambda)\csch(2\pi\lambda) 
+\frac{3}{4}(1+i\pi)\lambda I_{1}(2\pi\lambda) \csch(2\pi\lambda) \nn \\
&+&\frac{3}{4}\frac{\lambda^{2}}{\sinh(2\pi\lambda)} \int_{-1}^{+1}dr r\sqrt{1-r^{2}}\nn \\ 
& \times & \bigg\{(e^{-2\pi\lambda r}-e^{-2\pi\lambda})\Psi(i\lambda-i \lambda r)+(e^{2\pi\lambda}-e^{-2\pi \lambda r})\Psi(-i\lambda-i \lambda r)\bigg\}\Bigg].
\end{eqnarray}

Since the induced current is a physical observable, only its real part contributes. Therefore, taking the real part of Eq.~\eqref{regularizedcurrent} yields
\begin{eqnarray}\label{rewrite regularizedcurrent}
J& =& \frac{e}{2\pi^{2}}H^{2}\Omega^{-1}(\tau)\Bigg[-\frac{\pi}{4}(1+\lambda^{2})\coth(2\pi\lambda) \nn \\
&+&\frac{\pi}{4}(1-2\lambda^{2}) I_{0}(2\pi\lambda)\csch(2\pi\lambda) 
+\frac{3}{4}\lambda I_{1}(2\pi\lambda) \csch(2\pi\lambda) \nn \\
&+&\frac{3}{4}\frac{\lambda^{2}}{\sinh(2\pi\lambda)} \Re \int_{-1}^{+1}dr r\sqrt{1-r^{2}}\nn \\ 
& \times & \bigg\{(e^{-2\pi\lambda r}-e^{-2\pi\lambda})\Psi(i\lambda-i \lambda r)+(e^{2\pi\lambda}-e^{-2\pi \lambda r})\Psi(-i\lambda-i \lambda r)\bigg\}\Bigg],
\end{eqnarray}
where $\Re$ denotes the real part of the expression.

\section{Physical interpretation of the current} \label{sec:result}

\begin{figure}[ht]\centering
\includegraphics[width=\textwidth]{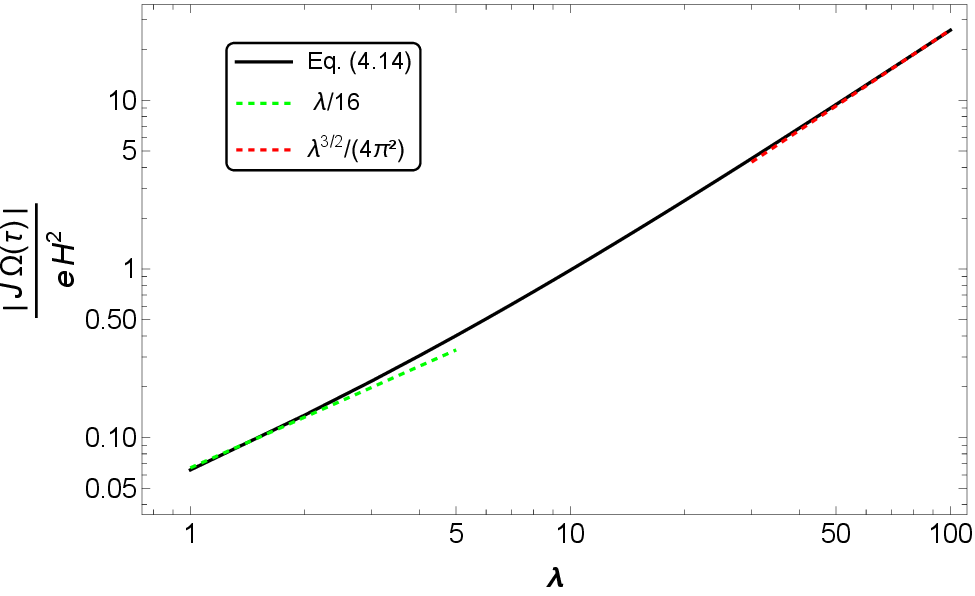}
\caption{ The renormalized induced current
$\left|J\Omega(\tau)\right|/(eH^2)$
as a function of the dimensionless parameter
$\lambda=eE/H^2$.
The black solid curve shows the evaluation of
Eq.~\eqref{rewrite regularizedcurrent}, where the integral over $r$ is computed numerically.
The green dashed and red dashed curves denote the weak-field and strong-field asymptotic behaviors,
$\left|J\Omega(\tau)\right|/(eH^2)\propto \lambda$ Eq.~\eqref{weakcurrent}
and
$\left|J\Omega(\tau)\right|/(eH^2)\propto \lambda^{3/2}$ Eq.~\eqref{analy},
respectively.
The numerical result smoothly interpolates between these two asymptotic regimes, approaching the linear weak-field behavior for $\lambda\ll1$ and the asymptotic $\lambda^{3/2}$ scaling in the strong-field regime.} 
\label{fig:1}
\end{figure}

In this section we analyze the physical behavior of the regularized in-vacuum expectation value of the induced current, given in Eq.~\eqref{rewrite regularizedcurrent}, with particular emphasis on its dependence on the external electric field. Our main result is presented in Fig.~\ref{fig:1}: the dependence of the normalized induced current on the electric field strength over a wide range of $\lambda$. We found that the induced current remains a continuous, monotonic, and analytic function of $\lambda$. The overall sign of the current depends on the sign convention adopted for the definition of $\lambda=eE/H^2$ (see also the discussion after Eq.~\eqref{lambda}). To gain further insight into the structure of the induced current, we examine its asymptotic behavior in the strong- and weak-field regimes. 
We decompose the renormalized current in Eq.~\eqref{rewrite regularizedcurrent} as
\begin{equation}
J=\frac{eH^2}{2\pi^2\Omega(\tau)}
\left( J_1+ J_2+ J_3\right),
\end{equation}
where
\begin{align}
& J_1 =-\frac{\pi}{4}(1+\lambda^{2})\coth(2\pi\lambda) 
+\frac{\pi}{4}(1-2\lambda^{2}) I_{0}(2\pi\lambda)\csch(2\pi\lambda), \nn\\
& J_2 =\frac{3}{4} \lambda I_{1}(2\pi\lambda) \csch(2\pi\lambda),\nn\\
& J_3 =\frac{3}{4}\frac{\lambda^{2}}{\sinh(2\pi\lambda)} \Re \int_{-1}^{+1}dr r\sqrt{1-r^{2}}\nn \\ 
& \times \bigg\{(e^{-2\pi\lambda r}-e^{-2\pi\lambda})\Psi(i\lambda-i \lambda r)+(e^{2\pi\lambda}-e^{-2\pi \lambda r})\Psi(-i\lambda-i \lambda r)\bigg\}.
\label{eq:decomp}
\end{align}

\begin{figure}[ht]\centering
\includegraphics[width=\textwidth]{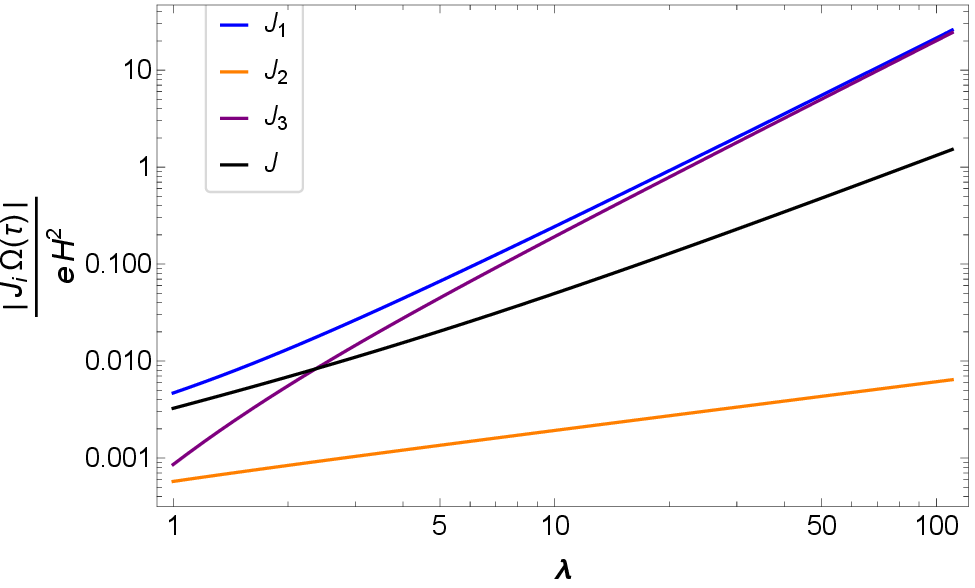}
\caption{Individual contributions $J_1$, $J_2$, and $J_3$, see Eqs.~\eqref{eq:decomp}, together with the resulting total current $J$.
The blue, orange, and purple curves correspond to $J_1$, $J_2$, and $J_3$,
respectively, while the black curve represents the total current $J$.
The dominant contributions $J_1$ and $J_3$ exhibit similar growth with
$\lambda$, but partially cancel each other, leading to a substantial
suppression of the total current in the intermediate-$\lambda$ regime.
The contribution $J_2$ remains subdominant throughout the entire range shown.}
\label{Fig:2}
\end{figure}

Figure~\ref{Fig:2} shows separately the contributions $J_1$, $J_2$, and $J_3$ entering Eq.~\eqref{rewrite regularizedcurrent}, together with their sum. The intermediate-$\lambda$ behaviour of the renormalized current can be traced to a partial cancellation between the dominant contributions $J_1$ and $J_3$. Although each contribution grows monotonically with $\lambda$, their combination leads to a substantial suppression of the total current in the intermediate regime. The contribution $J_2$ remains subdominant throughout the range shown. In contrast, the expected asymptotic scaling laws are recovered in both the weak- and strong-field limits.

\subsection{Asymptotic behaviors} \label{sec:behav}

For sufficiently large values of $\lambda$, the induced current exhibits the asymptotic scaling
$\left|J\right| \propto \lambda^{3/2}$,
in agreement with the strong-field analysis presented in Sec.~\ref{sec:strong}. This behaviour reflects the dominance of the Schwinger pair-production mechanism, whereby the produced particle--antiparticle pairs are efficiently separated by the electric field, giving rise to a macroscopic current. In contrast, in the weak-field regime, $\lambda \ll 1$, the induced current is strongly suppressed. The smooth crossover between these two asymptotic regimes demonstrates the consistency of the regularized expression for the induced current and provides a coherent description of vacuum polarization effects across the entire range of field strengths.

\subsubsection{Strong electric field} \label{sec:strong}
In the strong electric field regime, the asymptotic behavior of the induced current  $J$ \eqref{regularizedcurrent} is obtained by taking  $\lambda=\infty$. Using the saddle point integration method \cite{butler2007saddlepoint} we obtain the leading order terms as
\begin{eqnarray} \label{strongfield}
\frac{J }{e H^{2}}  &\simeq & -\frac{e}{4 \pi^{2}} \Omega^{-1}(\tau) \frac{|e E|^{\frac{1}{2}}}{H}E.
\end{eqnarray}
In the cases of $\ds_{2}$ \cite{frob2014schwinger}, $\ds_{3}$ \cite{bavarsad2016scalar} and $\ds_{4}$ \cite{kobayashi2014schwinger}, the authors showed that the current responds as $E^{1}$, $E^{\frac{3}{2}}$ and
$E^{2}$, respectively, in this regime.

In Ref.~\cite{bavarsad2016scalar}, it was shown that the semiclassical induced current in  $\mathrm{dS}$ spacetime scales as $J\propto E^{D/2}$ in the strong-field regime. Our result exhibits the same scaling behavior. Moreover, in the limit $\lambda \gg 1$, the fermionic current reduces to the bosonic result reported in Ref.~\cite{bavarsad2016scalar} (see Eqs.~(62) and (83) therein), as expected. This agreement reflects the fact that the strong-field regime effectively approaches the flat-spacetime Schwinger limit, where the electric field dominates over gravitational effects and the induced current becomes insensitive to the spin of the produced particles, leading to a universal behavior for both bosonic and fermionic cases.
We also note that the induced current remains positive for all values of $\lambda$ shown in Fig.~\ref{fig:1}. Therefore, plotting $|J|$ in Fig.~\ref{fig:1} does not affect the physical interpretation of the current direction.
It is also important to comment on the sign of the induced current in the strong-field regime. We find that the current flows along the direction of the background electric field, indicating a screening effect. This is consistent with the standard Schwinger mechanism picture, where particle–antiparticle pairs are produced by the electric field and subsequently accelerated in opposite directions, generating a current that tends to reduce the effective electric field.
Our analysis is performed assuming the Bunch--Davies vacuum, which provides a natural and well-defined
initial state in  $\mathrm{dS}$ spacetime. In the strong electric field limit, however, the induced current
is expected to be relatively insensitive to the specific choice of vacuum, since the electric field
dominates the dynamics over gravitational particle creation.

Finally, at sufficiently strong electric fields, the induced current may
lead to non-negligible backreaction effects, potentially modifying the
background electric field itself. Similar backreaction effects have been
investigated in  $\mathrm{dS}$ spacetime in several studies
\cite{stahl2016schwinger,sobol2018influence,sobol2019backreaction, kaushal2024backreaction}.
A fully self-consistent treatment including backreaction through
Maxwell's equations would be required to capture this effect, and we leave such an analysis for future work.

The numerical results are consistent with the analytically predicted strong-field scaling,
\begin{equation}\label{analy}
\frac{|J|}{eH^2}
\simeq
\frac{1}{4 \pi^{2}}\lambda^{3/2},
\qquad
\lambda \gg 1.
\end{equation}
As shown in Fig.~\ref{fig:1}, the numerical data closely follow the asymptotic curve in the large-$\lambda$ regime, confirming the predicted power-law behavior with exponent $3/2$. A fit of the strong-field regime yields a slope of 0.0262, reasonably close to the analytical expansion \eqref{strongfield}: $1/(4\pi^{2})=0.02533$.

\subsubsection{Weak electric field} \label{sec:weak}
In the weak-field regime, ($\lambda=\frac{eE}{H^2}\ll1$), the asymptotic behaviour of the induced current can be obtained from Eq.~\eqref{regularizedcurrent} by expanding the special functions around $\lambda=0$. Combining all contributions, the regularized current in the weak-field limit is given by
\begin{equation}
J=
-\frac{eH^2}{16}\,
\Omega^{-1}(\tau)\,
\lambda
+\mathcal{O}(\lambda^3).
\label{weakcurrent}
\end{equation}

Therefore, the induced current is linear in the electric-field strength in the weak-field limit. The cancellation of the apparent $1/\lambda$ singularities among the individual contributions ensures that the regularized current remains finite and analytic around $E=0$. Consequently, the current vanishes smoothly in the limit $E\to0$, demonstrating the absence of IR-HC in the massless fermionic case considered here.

The linear dependence on $E$ indicates that the vacuum response remains perturbative in this regime. Compared with the strong-field behavior, the induced current is strongly suppressed and can be interpreted as a small vacuum-polarization response generated by the external electric field. A fit of the weak-field regime yields a slope of 0.066, reasonably close to the Taylor expansion \eqref{weakcurrent}: $1/16=0.0625$.

\section{Conclusion}\label{sec:conclusion}
In this work, we have shown that the vacuum-induced fermionic current in three-dimensional $\mathrm{dS}$ spacetime exhibits distinct behaviors in the strong- and weak-field regimes. In the strong-field limit, the current is dominated by the Schwinger pair-production mechanism and follows the expected semiclassical scaling, whereas in the weak-field regime it is strongly suppressed and smoothly vanishes as the electric field approaches zero. These results were obtained from a finite expression for the renormalized induced current derived using adiabatic regularization in the Bunch--Davies vacuum.
Our analysis shows that, for sufficiently strong electric fields, the induced current is dominated by Schwinger pair-production. In this regime, the current exhibits the expected semiclassical scaling behavior and increases rapidly with the electric field strength. The sign of the current corresponds to a screening response, indicating that the produced charged particles tend to reduce the background electric field. This result is consistent with the physical picture of charged pairs accelerated by the external field and provides a nontrivial consistency check of the regularized expression.
In the weak-field regime, we found that the apparent singular contributions arising in the intermediate steps of the calculation cancel exactly after regularization. As a consequence, the physical current remains finite and analytic in the limit of a vanishing electric field. The induced current is strongly suppressed in this regime and can be interpreted as a small vacuum-polarization response generated by the external field. The smooth disappearance of the current as the electric field is switched off demonstrates the absence of infrared divergences in the physical observable.
A particularly important result of this work is the absence of IR-HC in the massless fermionic case. Unlike certain scalar-field models, where infrared effects may lead to an enhancement of the induced current, the fermionic current remains finite and decreases continuously as the electric field approaches zero. This behavior highlights the stabilizing role of fermionic statistics in  $\mathrm{dS}$ spacetime and suggests that the infrared properties of fermionic and scalar theories can be qualitatively different.
An additional feature of the renormalized current is that, for a fixed convention of the dimensionless parameter $\lambda=eE/H^2$, it remains continuous and monotonic throughout the parameter space. In particular, unlike the fermionic current in $\mathrm{dS}_4$, no sign change occurs.
Overall, our results show that the massless fermionic theory provides a consistent and well-behaved description of the Schwinger effect in  $\mathrm{dS}$ spacetime. The interplay between spacetime curvature and the external electric field gives rise to distinct strong- and weak-field regimes while preserving the regularity of the induced current throughout the parameter space.
Several interesting directions remain for future investigation. These include the study of massive fermions, time-dependent electromagnetic backgrounds, and a self-consistent treatment of the backreaction of the induced current on the background electric field. It would also be worthwhile to explore possible connections between induced fermionic currents in curved spacetime and anomaly-related phenomena in lower-dimensional QFTs.
The present massless analysis also provides a useful starting point for future investigations of massive fermions in $\mathrm{dS}_3$, where both parity-even Dirac mass and parity-odd topological (Haldane) mass terms may be present, leading to a richer structure of the induced current.

\section*{Authors' Contributions}
M.B. carried out all calculations, performed the numerical analysis, and prepared the manuscript. C.S. provided detailed feedback on the derivations, figures, and physical interpretation of the results, and contributed to improving the clarity of the presentation. All authors reviewed and approved the final manuscript.

\appendix
\section{Whittaker functions}
\label{app:whittaker}
In this Appendix, we have represented some useful relations and properties of Whittaker functions $\wwp(z)$ and $\wmp(z)$. More relations can be found in, e.g.~Ref.~\cite{frank2010nist}.\\
The Whittaker differential equation is
\begin{equation}\label{whittaker}
\frac{d^{2}}{dz^{2}}F(z)+\Big(-\frac{1}{4}+\frac{\kappa}{z}+\frac{\frac{1}{4}-\gamma^{2}}{z^{2}}\Big)F(z)=0,
\end{equation}
It has the two linearly independent solutions namely, $\wwp(z)$ and $\wmp(z)$. The needed connection formulas are
\begin{align}\label{ast}
\big(\w_{\kappa,\gamma}(z)\big)^{\ast} =&\w_{\kappa^{\ast},\gamma^{\ast}}(z^{\ast}),  
& \big(\m_{\kappa,\gamma}(z)\big)^{\ast} &=\m_{\kappa^{\ast},\gamma^{\ast}}(z^{\ast}),
\end{align}
\begin{align}\label{conection}
\w_{\kappa,\gamma}(z)&:=\w_{\kappa,-\gamma}(z), & \m_{\kappa,\gamma}(e^{\pm i\pi}z)&=\pm ie^{\pm i\pi\gamma}\m_{-\kappa,\gamma}(z).
\end{align}
The asymptotically expansion of the Whittaker functions as $|z|\rightarrow\infty$ are given by
\begin{align}
\label{win}
\wwp(z)&\sim e^{-\frac{z}{2}}z^{\kappa}, \\
\wmp(z)&\sim\frac{\Gamma(1+2\gamma)}{\Gamma(\frac{1}{2}+\gamma-\kappa)}
\,e^{\frac{z}{2}}z^{-\kappa} +\frac{\Gamma(1+2\gamma)}{\Gamma(\frac{1}{2}+\gamma+\kappa)} \,e^{-\frac{z}{2}\pm(\frac{1}{2}+\gamma-\kappa)\pi i}z^{\kappa}, \nn\\
\label{min}
-\frac{1}{2}\pi+\delta&\leq\pm{\rm ph}(z)\leq\frac{3}{2}\pi-\delta,
\end{align}
where $\delta$ is an arbitrary small positive constant. In the limit $|z|\rightarrow0$, asymptotically expansions are given by
\begin{align}
\label{mout}
&\wmp(z)\sim z^{\frac{1}{2}+\gamma}, \\
\label{wout}
&\wwp(z)\sim\frac{\Gamma(2\gamma)}{\Gamma(\frac{1}{2}+\gamma-\kappa)}
z^{\frac{1}{2}-\gamma}+\frac{\Gamma(-2\gamma)}{\Gamma(\frac{1}{2}-\gamma-\kappa)}
z^{\frac{1}{2}+\gamma}, & 0\leq\Re(\gamma)<\frac{1}{2}, &&\gamma\neq0.
\end{align}
Finally, some useful Wronskians are
\begin{eqnarray}
\label{wrww}
{\cal{W}}\Big\{\wwp(z),\w_{-\kappa,\gamma}(e^{\pm i\pi}z)\Big\}&=&e^{\mp i\pi\kappa}, \\ \label{wrmm}
{\cal{W}}\Big\{\wmp(z),\m_{\kappa,-\gamma}(z)\Big\}&=&-2\gamma, \\ \label{wrmw}
{\cal{W}}\Big\{\wwp(z),\wmp(z)\Big\}&=&\frac{\Gamma(1+2\gamma)}{\Gamma(\frac{1}{2}+\gamma-\kappa)}.
\end{eqnarray}
The recurrence relations used in this paper are
\begin{align}
\label{recurrenceW}
\w_{\kappa+\frac{1}{2},\gamma+\frac{1}{2}}(z)-\sqrt{z}\w_{\kappa,\gamma}(z)+(\kappa-\gamma-\frac{1}{2})\w_{\kappa-\frac{1}{2},\gamma+\frac{1}{2}}(z)=0,  \\
\w_{\kappa+\frac{1}{2},\gamma-\frac{1}{2}}(z)-\sqrt{z}\w_{\kappa,\gamma}(z)+(\kappa+\gamma-\frac{1}{2})\w_{\kappa-\frac{1}{2},\gamma-\frac{1}{2}}(z)=0, \\
(\kappa-\gamma-\frac{1}{2})\m_{\kappa-\frac{1}{2},\gamma+\frac{1}{2}}(z)+(1+2\gamma)\sqrt{z}\m_{\kappa,\gamma}(z)-(\kappa+\gamma+\frac{1}{2})\m_{\kappa+\frac{1}{2},\gamma+\frac{1}{2}}(z)=0, \\
2\gamma\m_{\kappa-\frac{1}{2},\gamma-\frac{1}{2}}(z)-2\gamma\m_{\kappa+\frac{1}{2},\gamma-\frac{1}{2}}(z)-\sqrt{z}\m_{\kappa,\gamma}(z)=0.
\end{align}
The differentiation formulas used in this paper are
\begin{align}
\label{differentiation}
\frac{d}{d z}\big(e^{-\frac{z}{2}}z^{k}\w_{\kappa,\gamma}(z)\big)=-e^{\frac{z}{2}}z^{1-\kappa}\w_{\kappa+1,\gamma}(z), \\
\frac{d}{d z} \big(e^{\frac{z}{2}}z^{-k}\w_{\kappa,\gamma}(z)\big)=e^{-\frac{z}{2}z^{1+\kappa}}(\frac{1}{2}-\kappa+\gamma)(\frac{1}{2}-\kappa-\gamma)\w_{\kappa-1,\gamma}(z), \\
\frac{d}{d z}\big(e^{-\frac{z}{2}}z^{k}\m_{\kappa,\gamma}(z)\big)=-e^{-\frac{z}{2}z^{1-\kappa}}(\frac{1}{2}+\kappa+\gamma)\m_{\kappa+1,\gamma}(z), \\
\frac{d}{d z} \big(e^{\frac{z}{2}}z^{-k}\m_{\kappa,\gamma}(z)\big)=e^{-\frac{z}{2}z^{1+\kappa}}(\frac{1}{2}-\kappa+\gamma)\m_{\kappa-1,\gamma}(z).
\end{align}
\section{Computation of the integral for the current}
\label{computation}
In this Appendix, the computation of the current integral ~(\ref{currentJP}) is presented. We follow the calculational procedure of Ref.~\cite{stahl2016fermionic} for a one-dimensional and Ref.~\cite{hayashinaka2016fermionic} for a three-dimensional momentum integral. We deal with the following integral:
\begin{eqnarray}\label{mathcalJ}
\mathcal{J}= \mathcal{J}_{1}-\mathcal{J}_{2},
\end{eqnarray}

where
\begin{eqnarray}
\mathcal{J}_{1} :&=& \int_{-1}^{+1}\frac{dr (1+r)}{\sqrt{1-r^{2}}} \int_{0}^{\Lambda}  dp\, p\, e^{-\pi \lambda r} 
 \big|W_{i\lambda r, i\lambda-\frac{1}{2}}(-2ip)\big|^{2},  \label{mathcalJ1} \\
\mathcal{J}_{2} &:=& \int_{-1}^{+1}\frac{dr (1-r)}{\sqrt{1-r^{2}}} \int_{0}^{\Lambda}  dp\, p\, e^{-\pi \lambda r} 
\big|W_{i\lambda r,i\lambda +\frac{1}{2}}(-2ip)\big|^{2}. \label{mathcalJ2}
\end{eqnarray}
Let us rewrite the Whittaker functions by using its Mellin-Barnes representation
\begin{equation}
\begin{split}\label{B4}
\w_{\kappa , \gamma}(z)&=e^{-\frac{z}{2}} \int_{-\infty}^{\infty} \frac{d s}{2\pi i}\frac{\Gamma(\frac{1}{2}+\gamma +s) \Gamma(\frac{1}{2}-\gamma +s)\Gamma(-\kappa -s)}{\Gamma(\frac{1}{2}+\gamma -\kappa)\Gamma(\frac{1}{2}-\gamma -\kappa)} z^{-s},\\
&|ph(z)| < \frac{3\pi}{2},\;\;\; \frac{1}{2}\pm \gamma -\kappa\neq 0,-1,-2,\cdots,
\end{split}
\end{equation}
where $ph$ is the phase and the contour of integration separates the poles of $\Gamma(\frac{1}{2}+\gamma+s) \Gamma(\frac{1}{2}-\gamma+s)$ from those of $\Gamma(-\kappa-s)$ \cite{frank2010nist}. By using the relations \eqref{ast} and \eqref{conection} , the integral \eqref{mathcalJ1}  can be rewritten as
\begin{eqnarray}\label{remathcalJ1}
\mathcal{J}_{1}& =& \int_{-1}^{+1}\frac{dr}{\sqrt{1-r^{2}}} C_{r}(1+r) 
\int_{-i\infty}^{i\infty}\frac{ds}{2\pi i} \Gamma(i \lambda+s) \Gamma(1-i\lambda+s)\Gamma(-i\lambda r-s) \nn \\
&\times &\int_{-i\infty}^{i\infty}\frac{dt}{2\pi i} \Gamma(-i\lambda+t) \Gamma(1+i\lambda+t)\Gamma(i\lambda r-t) e^{\frac{i\pi}{2}(s-t)} 2^{-s-t} \int_{0}^{\Lambda} dp p^{-s-t+1},
\end{eqnarray}
The definition of  $C_{r}$  is as follows:
\begin{align}\label{Cr}
 C_{r}=e^{-\pi \lambda r} \Big(\Gamma(1+i\lambda+i\lambda r)\Gamma(1-i\lambda-i\lambda r)\Gamma(i\lambda-i\lambda r)\Gamma(-i\lambda+i\lambda r)\Big)^{-1}.
\end{align}
In equation \eqref{remathcalJ1}, the integral over $p$ is evaluated as
\begin{equation}\label{intp}
\int_{0}^{\Lambda} p^{-s-t-1} \, dp = \frac{1}{- s - t+2}\, \Lambda^{- s - t+2}.
\end{equation}
In order for the result of the integral to vanish in the limit $\Lambda \to \infty$, as required, the integration contours for $s$ and $t$ must be chosen such that they satisfy the following condition:
\begin{equation} \label{cono}
\Re(s + t) < 2.
\end{equation}
To satisfy Eq.~\eqref{cono}, the integration contours for $s$ and $t$ are chosen such that:
\begin{equation} \label{cont}
\Re(s) < 1, \qquad \Re(t) < 1.
\end{equation}
Substituting the result of the integral~\eqref{intp} into equation~\eqref{remathcalJ1}, we obtain
\begin{align} \label{Jnew}
\mathcal{J}_{1}= C_r \int_{-i\infty}^{+i\infty} \frac{ds}{2\pi i}
\Gamma\left(i\lambda+s\right)
\Gamma\left(1-i\lambda+s\right)
\Gamma\left(-i\lambda r-s\right)
\int_{-i\infty}^{+i\infty} \frac{dt}{2\pi i}
\, f_r,s(t),
\end{align}
where, for fixed $s$ and $r$, the function $f_{r,s}(t)$ is defined as
\begin{align}\label{frst}
f_r,s(t) =\;&
\Gamma\left(-i\lambda+t\right)
\Gamma\left(1+i\lambda+t\right)
\Gamma\left(i\lambda r-t\right)\,
e^{\frac{i\pi}{2}(s-t)}
\frac{(2\Lambda)^{\,-s-t+2}}{4(-s-t+2)}
\end{align}

To evaluate the integral in Eq.~\eqref{Jnew}, we use the residue theorem. 
The poles of the integrand arise from the Gamma functions 
$\Gamma\left(-i\lambda+t\right)$, 
$\Gamma\left(1+i\lambda+t\right)$ 
and $\Gamma\left(i\lambda r -t\right)$. 
Thus, the poles of $f_{r,s}(t)$ are given by

\begin{equation} \label{conditionone}
 t_L = -n +i\lambda,  -n -i\lambda-1 \qquad
t_R = n+i\lambda ,
\end{equation}

where $n \in \mathbb{N}$.
Also, there is a simple pole arising from the denominator in \eqref{frst}, at
\begin{equation} \label{ind}
    t_{R} = 2- s.
\end{equation}
The poles $t_L$ arise from the Gamma functions 
$\Gamma\left(-i\lambda+t\right)$ and 
$\Gamma\left(1+i\lambda+t\right)$, 
which lie on the left-hand side of the integration contour. 
On the other hand, the poles $t_R$ originate from 
$\Gamma\left(i\lambda r-t\right)$, 
which lie on the right-hand side of the contour. 
We choose to close the integration contour in the right half-plane. 
Furthermore, we impose the following additional condition, together with condition~\eqref{conditionone}:
\begin{equation}\label{conditiontwo}
\Re(s) > -1.
\end{equation}

As a result, only the following poles contribute:
\begin{equation}\label{pole}
t_R = i\lambda r,\;  1+ i\lambda r,\; 2+ i\lambda r,\;  2 - s.
\end{equation}
In the limit $\Lambda \to \infty$, the non-vanishing residues remain. 
Therefore, the result of the integral over $t$ in Eq.~\eqref{Jnew} can be written as

\begin{equation}\label{It_sum}
\int_{-i\infty}^{+i\infty} \frac{dt}{2\pi i}\, f_r,s(t) =  h_{\mathrm{dep}(0)}(r,s)
+ h_{\mathrm{dep}(1)}(r,s)
+ h_{\mathrm{dep}(2)}(r,s)
+ h_{\mathrm{ind}}(r,s),
\end{equation}
where the residue at the pole $t_R =i\lambda r$ is given by
\begin{equation}\label{h0}
h_{\mathrm{dep}(0)}(r,s)
=  \Gamma\left(-i\lambda + i\lambda r\right)
\Gamma\left(1+i\lambda + i\lambda r\right)
\frac{(2\Lambda)^{\,2 - s - i\lambda r}}{4(2 - s - i\lambda r)}
\, e^{\frac{i\pi}{2}(s -i\lambda r)},
\end{equation}
the residue at the pole $t_R = 1+ i\lambda r$ is given by
\begin{equation}\label{h1}
h_{\mathrm{dep}(1)}(r,s)
= -\Gamma\left(1-i\lambda +i\lambda r\right)
\Gamma\left(2+i\lambda + i\lambda r\right)
\frac{(2\Lambda)^{\,1 - s -i\lambda r}}{4(1 - s - i\lambda r)}
\, e^{\frac{i\pi}{2}(-1+s -i\lambda r)},
\end{equation}
the residue at the pole $t_R = 2+i\lambda r$ is given by
\begin{equation}\label{h2}
h_{\mathrm{dep}(2)}(r,s)
=  \Gamma\left(2-i\lambda +i\lambda r\right)
\Gamma\left(3+i\lambda + i\lambda r\right)
\frac{(2\Lambda)^{\, - s - i\lambda r}}{8( -s - i\lambda r)}
\, e^{\frac{i\pi}{2}(-2+s -i\lambda r)},
\end{equation}
the residue at the pole $t_R = 2-s$ is given by
\begin{equation}\label{h7ind}
h_{\mathrm{ind}}(r,s)
= \Gamma\left(2-i\lambda -s\right)
\Gamma\left(3+i\lambda -s\right)
\Gamma\left(-2+i\lambda r +s\right)
\, \frac{e^{i\pi (-1+s)}}{4},
\end{equation}
Taking into account Eq.~\eqref{It_sum}, we rewrite Eq.~\eqref{remathcalJ1} as follows:
\begin{equation}  \label{Jsum}
\mathcal{J}_{1}= J_{\mathrm{dep}(0)} + J_{\mathrm{dep}(1)} + J_{\mathrm{dep}(2)}  + J_{\mathrm{ind}},
\end{equation}
where
\begin{equation}
J_{\mathrm{dep}(0)} = C_r \int_{-i\infty}^{+i\infty} \frac{ds}{2\pi i}
\, \Gamma\!\left(i\lambda+s\right)
\Gamma\!\left(1-i\lambda+s\right)
\Gamma(-i\lambda r-s)\,
h_{\mathrm{dep}(0)}(r,s),
\label{eq:Jdep0}
\end{equation}

\begin{equation}
J_{\mathrm{dep}(1)} = C_r \int_{-i\infty}^{+i\infty} \frac{ds}{2\pi i}
\, \Gamma\!\left(i\lambda+s\right)
\Gamma\!\left(1-i\lambda+s\right)
\Gamma(-i\lambda r-s)\,
h_{\mathrm{dep}(1)}(r,s),
\label{eq:Jdep1}
\end{equation}

\begin{equation}
J_{\mathrm{dep}(2)} = C_r \int_{-i\infty}^{+i\infty} \frac{ds}{2\pi i}
\, \Gamma\!\left(i\lambda+s\right)
\Gamma\!\left(1-i\lambda+s\right)
\Gamma(-i\lambda r-s)\,
h_{\mathrm{dep}(2)}(r,s),
\label{eq:Jdep2}
\end{equation}

\begin{equation}
J_{\mathrm{ind}} = C_r \int_{-i\infty}^{+i\infty} \frac{ds}{2\pi i}
\, \Gamma \left(i\lambda+s\right)
\Gamma \left(1-i\lambda+s\right)
\Gamma(-i\lambda r-s) h_{\mathrm{ind}}(r,s).
\label{eq:Jind}
\end{equation}

The poles of the integrand of $J_{\mathrm{dep}(0)}$ \eqref{eq:Jdep0} are given as follows:
\begin{equation}\label{poleJ}
s_L = -n -i\lambda, -n +i\lambda-1
\qquad
s_R = n -i\lambda r,
\qquad n \in \mathbb{N}
\end{equation}
The left poles $s_L$ are given by the poles of 
$\Gamma \left(i\lambda+s\right)$ and $\Gamma \left(1-i\lambda+s\right)$, 
which lie on the left side of the integration contour. 
 The right poles $s_R$ arise from the poles of 
$\Gamma\left(-i\lambda r- s\right)$, 
which lie on the right side of the integration contour and of the fraction on the right-hand side of Eq.~\eqref{h0}. 
We choose the contour such that it closes on the right half-plane.
Then, there are poles at
\begin{equation}
s_R = -i\lambda r,\; 1-i\lambda r,\; 2-i\lambda r.
\end{equation}
In the limit $\Lambda \to \infty$, the residues at these poles do not vanish. From the above, the result of the integral Eq.~\eqref{eq:Jdep0} is obtained as follows:

\begin{align} \label{JFdep0}
J_{\mathrm{dep}(0)}
= \frac{1}{8}
&\Big[
(1+i\lambda-i\lambda r)
(i\lambda-i\lambda r)
(2-i\lambda-i\lambda r)
(1-i\lambda-i\lambda r)
\Big]
\nonumber\\
&\times
\Bigg\{\Psi(2+i\lambda-i\lambda r)
+\Psi(3-i\lambda-i\lambda r)
-\frac{3}{2}
+\gamma_E
+\frac{i\pi}{2}
-\ln(2\Lambda)\Bigg\}
\nonumber\\
&\qquad
-\frac{i}{2}
(i\lambda-i\lambda r)
(1-i\lambda-i\lambda r)\Lambda
+\frac{1}{2}\Lambda^2,
\end{align}
where $\gamma_E$ is the Euler--Mascheroni constant and $\Psi$ is the digamma function.
To obtain the result of the integrals $J_{\text{dep(1)}}$  and $J_{\text{dep(2)}}$, which are related to ~Eq.~\eqref{eq:Jdep1}--Eq.~\eqref{eq:Jdep2}, we proceed in a similar manner as for the integral $I_{\text{dep(0)}}$. Therefore, we have:
\begin{align} \label{JFdep1}
J_{\mathrm{dep}(1)}
=
-\frac{1}{4}
&( -i\lambda+i\lambda r )
( 1+i\lambda+i\lambda r )
( i\lambda-i\lambda r )
( 1-i\lambda-i\lambda r )
\nonumber\\
&\times
\Bigg\{
\Psi(1+i\lambda-i\lambda r)
+\Psi(2-i\lambda-i\lambda r)
-1+\gamma_E
+\frac{i\pi}{2}
-\ln(2\Lambda)
\Bigg\}
\nonumber\\
&\qquad
+\frac{i}{2}
(-i\lambda+i\lambda r)
(1+i\lambda+i\lambda r)\Lambda,
\end{align}
\begin{align} \label{JFdep2}
J_{\mathrm{dep}(2)}
=
\frac{1}{8}
&(1-i\lambda+i\lambda r)
(-i\lambda+i\lambda r)
(2+i\lambda+i\lambda r)
(1+i\lambda+i\lambda r)
\nonumber\\
&\times
\Bigg\{
\Psi(i\lambda-i\lambda r)
+\Psi(1-i\lambda-i\lambda r)
+\gamma_E
+\frac{i\pi}{2}
-\ln(2\Lambda)
\Bigg\}.
\end{align}
Using Eq.~\eqref{eq:Jind} and the property of the Gamma function 
\begin{align}
\label{eq:gamma2sin}
\Gamma(z)\Gamma(1-z)
&= \frac{\pi}{\sin(\pi z)}, \qquad z \notin \mathbb{Z},
\end{align}
 the integral $J_{\text{ind}}$ can be rewritten as follows:
\begin{align}\label{JFind}
J^{(\mathrm{ind})}
={}&
C_r \frac{\pi^3}{4}
\int_{-i\infty}^{+i\infty}
\frac{ds}{2\pi i}\,
\frac{
(1-i\lambda-s)
(2+i\lambda-s)
(1+i\lambda-s)
(i\lambda-s)\,
e^{i\pi s}
}{
\sin\pi(s+i\lambda)
\sin\pi(s-i\lambda)
}
\nonumber\\[2mm]
&\times
\frac{1}{
(2-s-i\lambda r)
(1-s-i\lambda r)
(-s-i\lambda r)
\sin\pi(s+i\lambda r)
}.
\end{align}

The poles of the integrand~\eqref{JFind} are given as follows ($n \in \mathbb{N})$:
\begin{align}
s_R &=  n - i\lambda , \qquad
s_L = - n - i\lambda , \label{eq:poles1} \\
s_R &=  n + i\lambda, \qquad
s_L =  - n +i\lambda-1,\label{eq:poles2} \\
s_R &=  n - i\lambda r, \qquad
s_L = - n - i\lambda r-1.\label{eq:poles3}
\end{align}
 The poles of $\sin\pi(s + i\lambda )$ are in Eq.~\eqref{eq:poles1}, the poles of $\sin\pi(s -  i\lambda)$ are in Eq.~\eqref{eq:poles2}, and the poles of $\sin\pi(s + i\lambda r)$ are in Eq.~\eqref{eq:poles3}.
We consider the following expression:
\begin{equation}
U_{r}(s)=
\frac{(s-i\lambda-2)(s+i\lambda-1)(s-i\lambda-1)(s-i\lambda)}
     {(s+i\lambda r-2)(s+i\lambda r-1)(s+i\lambda r)},
\end{equation}
which can be rewritten as follows:
\begin{equation}
U_{r}(s) = -\frac{6r\lambda^{2}(1+r)}{s+i\lambda r}
+g_{r}(s)-g_{r}(s-1),
\label{eq:B3_gr}
\end{equation}
where $ g_r(s)$ is given by
\begin{equation}
 g_r(s) =
-\frac{b(r)+c(r)}{s+i\lambda r}
-\frac{c(r)}{s+i\lambda r-1}
+\frac{1}{2}
\left(s^2+s\right)
+\left(-2i\lambda-3i\lambda r-1\right)(s+1),
\end{equation}
with
\begin{equation}
b(r)=-
\left(
-\lambda^{2}
+r^{2}\lambda^{2}
-\lambda^{4}
-2r\lambda^{4}
+2r^{3}\lambda^{4}
+r^{4}\lambda^{4}
\right).
\end{equation}
\begin{align}
c(r)
&=
\frac{1}{2}\Bigl(
-2i\lambda
-2ir\lambda
-\lambda^{2}
-6r\lambda^{2}
-5r^{2}\lambda^{2}
-2i\lambda^{3}
\\
&\qquad
+6ir^{2}\lambda^{3}
+4ir^{3}\lambda^{3}
-\lambda^{4}
-2r\lambda^{4}
+2r^{3}\lambda^{4}
+r^{4}\lambda^{4}
\Bigr).
\end{align}
By substituting Eq.~\eqref{eq:B3_gr} into Eq.~\eqref{JFind}, we obtain
\begin{equation}\label{Jindfinal}
\begin{aligned}
J^{(\mathrm{ind})}
=&
\left(-\frac{\pi^{3}}{4}\right)
C_r
\int_{-i\infty}^{+i\infty}
\frac{ds}{2\pi i}\,
\frac{e^{i\pi s}}
{\sin\!\pi(s+i\lambda)
 \sin\!\pi(s-i\lambda)
 \sin\!\pi(s+i\lambda r)}
\\
&\times
\left[
-\frac{6r\lambda^{2}(1+r)}
{s+i\lambda r}
+g_r (s)
-g_r (s-1)
\right].
\end{aligned}
\end{equation}
For convenience in evaluating $J^{(\mathrm{ind})}$, we rewrite it as

\begin{equation}\label{Jinf}
J^{(\mathrm{ind})}= J^{(\mathrm{indu})} + J^{(\mathrm{indf})},
\end{equation}

where $J^{\mathrm{indu}}$ and $J^{\mathrm{indf}}$ are defined by
\begin{align} \label{Jindu}
J^{(\mathrm{indu})}
&=
\left(-\frac{\pi^{3}}{4}\right)
C_{r}
\int_{-i\infty}^{+i\infty}
\frac{ds}{2\pi i}\,
\frac{e^{i\pi s}}
{\sin\!\pi(s+i\lambda)
 \sin\!\pi(s-i\lambda)
 \sin\!\pi(s+i\lambda r)}
\left[
g_{r}(s)-g_{r}(s-1)
\right],
\end{align}

\begin{align}\label{Jindf}
J^{(\mathrm{indf})}
&=
\left(-\frac{\pi^{3}}{4}\right)
C_{r}
\int_{-i\infty}^{+i\infty}
\frac{ds}{2\pi i}\,
\frac{e^{i\pi s}}
{\sin\pi(s+i\lambda)
 \sin\pi(s-i\lambda)
 \sin(\pi(s+i\lambda r)}
\left(
\frac{-6r\lambda^{2}(1+r)}
{s+i\lambda r}
\right).
\end{align}
The poles of the integrand~\eqref{Jindu} are given as follows ($n \in \mathbb{N}$):
\begin{align}
s_R &=  n - i\lambda, \qquad
s_L = - n - i\lambda, \label{eq:poleslr1} \\
s_R &= n + i\lambda, \qquad
s_L =-n + i\lambda-1, \label{eq:poleslr2} \\
s_R &= n-i\lambda r, \qquad
s_L = -n -i\lambda r-1. \label{eq:poleslr3}
\end{align}
To evaluate the integral~\eqref{Jindu}, we close the contour of integration in the left half-plane. Thus, it can be concluded that the contribution of the poles of $g_r(s)$ at $s=n$ is canceled by the contribution of the poles of $g_r(s-1)$ at $s=n+1$. Therefore, only the contributions of the following poles remain in the term $g_r(s)$:
\begin{equation} \label{poleleft}
    s_L = -i \lambda,\quad i\lambda-1,\quad -i\lambda r-1,
\end{equation}
hence, the integral~~\eqref{Jindu} simplifies as follows
\begin{align} \label{Jindusim}
J^{(\mathrm{indu})}
&=
\left(-\frac{\pi^{3}}{4}\right)
C_{r}
\int_{-i\infty}^{+i\infty}
\frac{ds}{2\pi i}\,
\frac{e^{i\pi s}}
{\sin\!\pi(s+i\lambda)
 \sin\!\pi(s-i\lambda)
 \sin\!\pi(s+i\lambda r)}
g_{r}(s).
\end{align}
Using Eq.~\eqref{eq:gamma2sin}, $C_{r}$ in Eq.~\eqref{Cr} can be rewritten as follows:
\begin{equation}\label{Crewrite}
C_{r}
=
-\frac{
e^{-\pi\lambda r}
\left(i\lambda-i\lambda r\right)
\sin \pi(i\lambda-i\lambda r)
\sin\pi(i\lambda+i\lambda r)
}{
\pi^{2}\left(i\lambda+i\lambda r\right)
}.
\end{equation}
Using the residue theorem for the poles of Eq.~\eqref{poleleft}, we integrate Eq.~\eqref{Jindusim}:
\begin{align}\label{Jindsum}
J^{(\mathrm{indu})}
={}&
\frac{i}{8}\,
\csc(2i\pi\lambda)\,
e^{-2\pi\lambda r}\,
\Big(\frac{1-r}{1+r}\Big)
\left[
g_r(i\lambda-1)-g_r(-i\lambda)
\right]
\nonumber\\
&+\frac{i}{8}\,
\cot(2i\pi\lambda)\,
\Big(\frac{1- r}{1+ r}\Big)
\left[
g_r(-i\lambda)-g_r(i\lambda-1)
\right]
\nonumber\\
&+\frac{1}{8}\,
\Big(\frac{1-r}{1+r}\Big)
\left[
g_r(-i\lambda)+g_r(i\lambda-1)
\right]
\nonumber\\
&-\frac{1}{4}\,
\Big(\frac{1-r}{1+r}\Big)
g_r(-i\lambda r-1)
\end{align}
To evaluate the integral in Eq.~\eqref{Jindf}, we close the contour in the left half-plane. Therefore, the simple poles $s_L$, given by Eqs.~\eqref{eq:poleslr1}--\eqref{eq:poleslr3}, lie inside the contour. Applying the residue theorem, we obtain
\begin{align}\label{Jindfsigma}
J^{(\mathrm{indf})}
&=
\left(-\frac{\pi^{3}}{4}\right)
C_{r}
\Bigg[-\frac{e^{\pi\lambda}}
{\pi\sin(2\pi i\lambda)\sin\pi(i\lambda-i\lambda r)}
\sum_{n=0}^{\infty}
\frac{1}{n+i\lambda-i\lambda r}
\nonumber\\
&-\frac{e^{-\pi\lambda}}
{\pi\sin(2\pi i\lambda)\sin\pi(i\lambda+i\lambda r)}
\sum_{n=0}^{\infty}
\frac{1}{\,n+1-i\lambda-i\lambda r\,}
\nonumber\\
&-\frac{e^{\pi\lambda r}}
{\pi\sin\pi(i\lambda-i\lambda r)
 \sin\pi(-i\lambda-i\lambda r)}
\sum_{n=0}^{\infty}
\frac{1}{n+1}\Bigg]\Big(-6r\lambda^{2}(1+r)\Big).
\end{align}
Using the definition of the Hurwitz zeta function
\begin{equation}\label{eq:HurwitzZeta}
\zeta(c,z)=\sum_{n=0}^{\infty}\frac{1}{(n+z)^c},
\qquad
\Re(c)>1,\quad
z \notin \mathbb{Z}^-.
\end{equation}
Taking Eqs.~\eqref{eq:HurwitzZeta} and \eqref{Crewrite} into account, Eq.~\eqref{Jindfsigma} can be written as
\begin{align} \label{JHur}
J^{(\mathrm{indf})}
&=
\left(-\frac{1}{4}\right)
\Big(\frac{1-r}{1+r}\Big)
\Bigg[
\frac{\left(e^{-2\pi\lambda r}-e^{2\pi\lambda}\right)}
{2i\sin(2\pi i\lambda)}
\,\zeta\!\left(1,i\lambda-i\lambda r\right)
\nonumber\\
&+
\frac{\left(e^{-2\pi\lambda}-e^{-2\pi\lambda r}\right)}
{2i\sin(2\pi i\lambda)}
\,\zeta\!\left(1,-i\lambda-i\lambda r+1\right)
-
\zeta(1,1)
\Bigg]\Big(-6r\lambda^{2}(1+r)\Big).
\end{align}
The expansion of the Hurwitz zeta function around $c=1$ is given by \cite{frank2010nist}
\begin{equation}
\zeta(c,z)=\frac{1}{c-1}-\Psi(z)+\mathcal{O}(c-1),
\label{eq:zetaexp}
\end{equation}
Eq.~\eqref{JHur} can thus be written
\begin{equation}\label{Jindfsum}
\begin{aligned}
J^{(\mathrm{indf})}=
\left(-\frac{3}{2}\right)(1-r)\,r\lambda^{2}
\Biggl[
&\frac{1}{2i\sin(2i\pi\lambda)}
\Bigl\{
\left(e^{-2\pi\lambda r}-e^{2\pi\lambda}\right)
\Psi\!\left(i\lambda-i\lambda r\right)
\\
&\qquad
+\left(e^{-2\pi\lambda}-e^{-2\pi\lambda r}\right)
\Psi\!\left(-i\lambda-i\lambda r +1\right)
\Bigr\}
+\delta_{E}
\Biggr].
\end{aligned}
\end{equation}

Substituting Eqs.~\eqref{JFdep0}, \eqref{JFdep1}, \eqref{JFdep2}, \eqref{Jinf}, \eqref{Jindsum}, and \eqref{Jindfsum} into Eq.~\eqref{Jsum} and simplifying, we obtain $\mathcal{J}_{1}$. 
Replacing $\gamma$ with $-\gamma$ in Eq.~\eqref{mathcalJ1} and simplifying, we obtain the corresponding expression for $\mathcal{J}_{2}$.
Substituting Eqs.~\eqref{mathcalJ1} and \eqref{mathcalJ2} into Eq.~\eqref{mathcalJ} and simplifying, we obtain
\begin{eqnarray}\label{mathcalJfinal}
\mathcal{J} &=&\pi \lambda\Lambda-\frac{3}{4}i\pi\lambda-\frac{\pi}{2}(1+\lambda^{2})\coth(2\pi\lambda) \nn \\
&+&\frac{\pi}{2}(1-2\lambda^{2}) I_{0}(2\pi\lambda)\csch(2\pi\lambda) 
+\frac{3}{2}(1+i\pi)\lambda I_{1}(2\pi\lambda) \csch(2\pi\lambda) \nn \\
&+&\frac{3}{2}\frac{\lambda^{2}}{\sinh(2\pi\lambda)} \int_{-1}^{+1}dr r\sqrt{1-r^{2}}\nn \\ 
& \times & \bigg\{(e^{-2\pi\lambda r}-e^{-2\pi\lambda})\Psi(i\lambda-i \lambda r)+(e^{2\pi\lambda}-e^{-2\pi \lambda r})\Psi(-i\lambda-i \lambda r)\bigg\}.
\end{eqnarray}

\bibliographystyle{JHEP}
\bibliography{bib}

\end{document}